\documentclass{ws-rv975x65}
\usepackage{ws-rv-van}             

\newcommand{\vr} { \mathbf {r}}
\newcommand{\vD} { \mathbf {D}}
\newcommand{\vj} { \mathbf {j}}
\newcommand{\vA} { \mathbf {A}}
\newcommand{\sex}[1]{\langle #1 \rangle}

\def\der#1#2{\frac{\partial #1}{\partial #2}}

\def \be{\begin{equation}}
\def \ee{\end{equation}}
\def \ba{\begin{array}}
\def \ea{\end{array}}
\def \bea{\begin{eqnarray}}
\def \eea{\end{eqnarray}}

\def \l{\left}
\def \r{\right}
\def \rr{\right}

\def \H{{\cal{H}}}
\def \L{{\cal{L}}}

\def \bE{{\bf E}}
\def \bj{{\bf j}}

\def \br{{\bf r}}

\def \intt{\int\limits}

\def \summ{\sum\limits}
\def\intt{\int\limits}

\begin{document}

\chapter{Resistance in Superconductors}

\author{Bertrand I. Halperin$^a$, Gil Refael$^b$, Eugene Demler$^a$
}

\address{$^a$Physics Department, Harvard University, Cambridge MA 02138\\
$^b$Department of Physics, California Institute of Technology, Pasadena, CA 91125
}
 
\begin{abstract} 
In this pedagogical review, we discuss how electrical resistance can arise
in superconductors. Starting with the idea of the superconducting order
parameter as a condensate wave function, we introduce vortices as
topological excitations with quantized phase winding, and we
show how phase slips occur when vortices cross the sample. Superconductors
exhibit non-zero electrical resistance under circumstances where phase
slips occur at a finite rate.  For one-dimensional superconductors or
Josephson junctions, phase slips can occur at isolated points in
space-time.  Phase slip rates may be controlled by thermal activation over
a free-energy barrier, or in some circumstances, at low temperatures, by
quantum tunneling through a barrier. We present an overview of several
phenomena involving vortices that have direct implications for the
electrical resistance of superconductors, including the
Berezinskii-Kosterlitz-Thouless transition for vortex-proliferation in
thin films, and the effects of  vortex pinning in  bulk type II
superconductors on the non-linear resistivity of these materials in an
applied magnetic field. We discuss how  quantum fluctuations can cause
phase slips  and review  the non-trivial role of dissipation on such fluctuations.  We present a basic picture of the superconductor-to-insulator quantum phase transitions in films, wires, and Josephson
junctions. We point out related problems in superfluid
helium films and systems of ultra-cold trapped atoms. 
 While our emphasis is on theoretical concepts, we also briefly
describe experimental results, and we underline some of the open questions.
 \end{abstract}

\body

\section{Introduction}\label{sec1.1}

The ability of a  wire to carry an electrical current with no apparent dissipation 
is doubtless the most dramatic property of the superconducting state.
Under favorable conditions, the electrical resistance of a
superconducting wire can be very low indeed. 
Mathematical models predict
lifetimes that far exceed the age of the universe for sufficiently
thick wires under appropriate conditions. In one experiment, a
superconducting ring was observed to carry a persistent current for
more than a year without measurable decay, with an  upper bound for
the decay rate of a part in 10$^5$ in the course of a year\cite{TinkhamBook}.    
However, in other
circumstances, as for sufficiently thin wires or films, or in the
presence of penetrating strong magnetic fields,  non-zero resistances
are observed.  Over the past fifty years, a great deal of theoretical
and experimental effort has been devoted to obtaining  better
qualitative and quantitative understanding of how this resistivity
arises, and how  superconductivity breaks down, in the variety of
possible situations.

Let us first recall how superconductors can exhibit negligible resistance in favorable situations.  
In a paper that  appeared seven years before  BCS,   Ginzburg and
Landau (GL) proposed that a superconductor should be characterized by
a complex-valued function of position,   $\Psi( \vr  )$, 
referred to as the {\em superconducting order parameter}. They  proposed, further, 
that one could define a free-energy functional $F$, which depends on
$\Psi(\vr)$, and which would be minimized, in the case of a
homogeneous superconductor with no external magnetic field, at
temperatures below the critical temperature $T_c$, by an order
parameter $\Psi(\vr)$ whose magnitude was a constant
$\Psi_0 > 0$, determined by the temperature $T$. In order to minimize the free energy, the phase of the complex
order parameter should be independent of position, but the value of the free energy would be
independent of the choice of this constant overall phase.

Following BCS, we  now understand that the microscopic origin of the GL order parameter is the  condensate wave function for Cooper pairs. Except for an arbitrary normalization constant, this is equal to   the anomalous expectation value 
$\sex {\psi_{\uparrow} (\vr) \psi_{\downarrow} (\vr)  }$, where
$\psi_{\uparrow} (\vr)$ and $ \psi_{\downarrow} (\vr) $ are the
annihilation operators for an electron at position $\vr$, with spin up
and spin down respectively. The existence of this non-zero expectation
value signifies that the superconducting state has a broken symmetry,
namely the $U(1)$ symmetry, commonly referred to as gauge symmetry,
associated with the  conservation of charge or electron
number. 



The GL assumption
enables us to understand the phenomenon of persistent
currents. Consider a superconducting ring in a situation where the magnitude of
the  order parameter is a constant,  but where the phase varies around
the ring.  Since the order parameter $\Psi$ must be single-valued at
any given position,  the net phase change around the ring must be an
integral multiple, $n$, of $2 \pi$.  A state with non-zero winding
number $n$ will have an {\em excess}  free energy,  proportional to
$n^2$ for moderate values of $n$.  As we shall see below, it  will also carry an
electrical current proportional to $n$.
One finds that in order for the initial state to decay to a state with a smaller winding
number, and hence, a smaller value of the free energy  and a smaller value of the
current, it is necessary for the system to pass through an
intermediate configuration of $\Psi(\vr)$ where the free energy is
larger  than the free
energy of the initial current carrying state, by an amount $\Delta F $.  Because $\Psi(\vr)$ is
the wave function for a large number of Cooper pairs, the free energy
barrier $\Delta F$ can be very large compared to the thermal energy
$k_{\rm B} T$, and the probability to be carried over the barrier by a
thermal fluctuation can be exceedingly small.  On the other hand, by
adjusting parameters, including the temperature and the dimensions of
the system, one may reduce $\Delta F$ to the point where transitions
occur at a small but measurable rate.

Quite generally, if we know the minimum value of the free energy barrier $\Delta F$ separating a state with winding number $n$ from a state with $n-1$, we may estimate a thermally-activated rate of transitions between the two states as 
\begin{equation}
\eta = \Omega \, e^{- \Delta F / T} \, ,
\label{activation}
\end{equation}
where  the prefactor $\Omega$ depends on details of the system, but is
generally a less-rapidly varying function of the parameters than the
exponential factor, in the region of interest.  (We  measure temperature in energy units, so that  $k_{\rm B}=1$.)   The most
important step, therefore,  in estimating the decay rate of persistent
current in a superconducting loop is to identify the process with
smallest  barrier and to calculate $\Delta F$ as accurately as possible.  After that, one must estimate the associated
prefactor $\Omega$.

\begin{center}
\begin{figure}
\includegraphics[width=10cm]{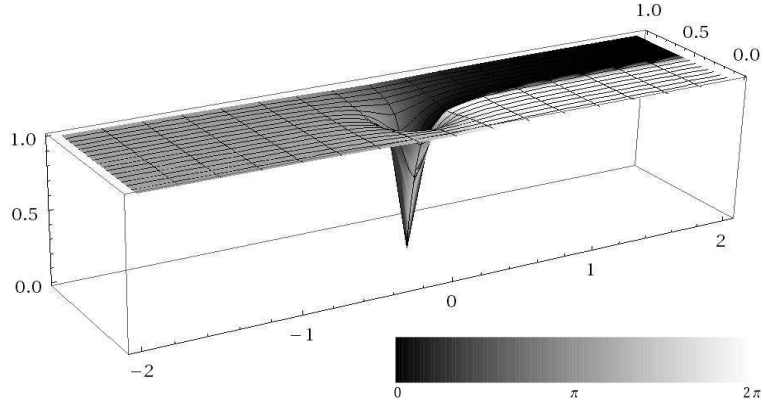}
\caption{An illustration of a vortex in a superconducting strip. The
  height of the function corresponds to the magnitude of the order
  parameter  $\Psi(\br)$   , and the greyscale to its phase $\phi(\br)$, mod $ 2\pi$. The greyscale discontinuity along the line $y=0.5$, for $x >0$, does not represent a discontinuity in $\Psi$, because $e^{i \phi}$ is continuous.   The coherence length
 used  here is $\xi=0.1$  . \label{vortexfig}}
\end{figure}
\end{center}

The most favorable path for a change in winding number, and the corresponding value of  $\Delta F$, can be quite different in different geometries, as we shall discuss below. However, there are some general considerations. The winding number $n$ may be defined as $(2 \pi)^{-1} $ times
the integral of the phase gradient $   \nabla \phi$ along a closed path around the ring, embedded in the superconductor.  
If we assume that $\Psi(\vr)$ varies  continuously as a function of
position and time, however,   the only way that the accumulated phase
can jump discontinuously at some time $t$,  is if $\Psi$ vanishes at
some point on the path. Since $\Psi$ has both a real and imaginary
part, the locus of points $(\vr,t)$ with $\Psi=0$ will generically
form a set of co-dimension 2.  In the case of a two-dimensional
superconductor, at a fixed time $t$, zeroes of $\Psi$ should occur at
isolated points, known as {\em vortex points}, which are described as
positive or negative, depending on whether the phase changes by $\pm 2
\pi$ as one winds around the vortex in a counterclockwise sense.  In
three dimensions, the vortex core, where $\Psi = 0$,  becomes a  line segment ending at the
boundaries of the  superconductor, or possibly  a closed loop,
embedded within the material. For a given location of the vortex core,
the state of minimum free energy will  generally occur
when the magnitude of $\Psi$  returns to its equilibrium value over a
healing distance  comparable to the BCS coherence length
$\xi(T)$. The structure of a vortex is depicted in Fig. \ref{vortexfig}. 

Consider a superconducting annulus
with a given initial phase-winding number
$n>0$. One way to change $n$ with the smallest
barrier $\Delta F$ would be to introduce a single positive vortex   at
the outer edge of the annulus, and move it across the ring until it
disappears at the inner edge.  Alternatively, we could move a negative
vortex from the inner edge to the outer.  In either case, any closed
curve around the ring would have been cut once by the vortex, and the
winding number would have changed by -1. The free energy cost for
introducing a vortex into the film will depend on the details of the
material, but will clearly increase with
increasing thickness of the film. The free energy barrier can therefore be made arbitrarily large by going to a wire of sufficient thickness.

In order to go further, it is helpful to  make some assumptions about the form of the free energy functional $F(\Psi)$. The specific  form proposed by Ginzburg and Landau  may be written as
\begin{equation}
F = \int d^3 \vr  \left[ - \frac{\alpha} {2} |\Psi|^2 + \frac{\beta}{4} | \Psi|^4 + \frac {\gamma}{2}
\left|  \left(  \nabla - \frac{2ie }{\hbar c} \vA \right) \Psi  \right| ^2   \right] \, ,
\label{GL}
\end{equation}
where $\vA (\vr)$ is the electromagnetic vector potential (We use Gaussian cgs units throughout this chapter).   The parameters $\alpha, \beta, \gamma$, are all positive below $T_c$, but $\alpha$ is assumed to vanish linearly as $T \to T_c$.  If there is no magnetic field present, 
the free energy is minimized when $|\Psi| = (\alpha/\beta) ^{1/2}$, and the resulting free energy reduction per unit volume is  $f_0 =(\alpha^2/4 \beta)$.
The parameters of (\ref{GL}) result in a coherence length $\xi$ and a  magnetic penetration depth $\lambda$ given by 
\begin{equation}
\xi = (\gamma/\alpha)^{1/2} \, , \, \, \, \, \,  \lambda = \kappa \xi \, , \, \, \, \, \,
\kappa^2  \equiv   \frac { \hbar^2  c^2  \beta  }{16 \pi e^2 \gamma} \, .
\end{equation}
We shall be interested primarily  in ``type II'' superconductors, where  $\kappa \gg 1$.  The free energy cost for creating a vortex is typically of order $f_0 \xi^2 L_v$, where $L_v$ is the length of the vortex.

The free energy (\ref {GL}) implies that in a state of  stable or metastable equilibrium, there will be an electrical current (``supercurrent''), whose density is given by  
\begin{equation}
\vj_s(\vr) = - c  \frac{ \delta F}{\delta \vA (\vr)} =
 \frac{2e \gamma}{\hbar}  | \Psi |^2  \left( \nabla \phi  - \frac{2e}{\hbar c} \vA   \right) \, ,
 \label{js}
\end{equation}
where $\phi(\vr)$ is the phase of $\Psi(\vr)$.
Although the specific forms (\ref{GL}) and (\ref{js}) are quantitatively correct only close to $T_c$, they are qualitatively correct more generally, and they will serve well for our purposes.   The BCS theory gives a prediction for the Ginzburg-Landau parameters in terms of  microscopic parameters of the material, such as the electron density of states,  the effective mass, the mean-free-path, and the electron-phonon interaction strength. 
 
The decay rate of a persistent current can be changed dramatically if
there is an applied magnetic field exceeding the threshold $H_{c1}=\frac{hc/2e}{4\pi\lambda^2}\ln\kappa$ for penetration into the superconductor.  In this case one finds that there are vortices present even in thermal equilibrium, with a vortex density $n_v$ proportional to the magnetic field: 
\begin{equation}
n_v = B / \Phi_0, ,
\end{equation}
where $\Phi_0 \equiv h c / 2e$ is the {\it superconducting flux quantum}.   Although the vortices will ideally form a triangular Abrikosov lattice, in practice the lattice is usually distorted and the vortices tend to be pinned by various types of inhomogeneities in the 
material.  In this case, since vortices are already present, the free
energy barrier for relaxation of a supercurrent at low temperatures
will be the activation energy necessary for vortices to become
unpinned so they can move across the current-carrying sample.

An analysis of the decay rate for persistent current in a ring can be extended directly to the onset of resistance in an open superconducting wire carrying a current between two contacts.  
Because $\Psi(\vr)$ represents a wave function for Cooper pairs of charge $2e$, there is a commutation relation between the phase $\phi$ and the total electron number $N$, given by 
\begin {equation}
[\phi,\, N ] =  2i  \, .
\label{commrel}
\end{equation}
Then, because electrical charge is conserved, it can be shown that in a state of local equilibrium, at any temperature $T$, the phase $\phi$ must evolve in time according to the Josephson relation \begin{equation}
d\phi/dt = 2e V / \hbar  \, ,
\label{josephson}
\end{equation}
where $V$ is the electrochemical potential ({\it i.e.}, the voltage) at the point $\vr$. 
[Equivalently, (\ref{josephson}) may be understood as a consequence of gauge invariance.]
If  the two ends of a superconducting wire were each  in local equilibrium, with their  voltages  differing by an amount $\Delta V \neq 0$, and if no vortices or phase slips were allowed to cross the intervening wire, the phase difference between the two ends of the wire would increase linearly in time, and the resulting supercurrent would increase without limit. In order to reach a steady state with a constant current, there must be a net flow of vortices or phase slips across the wire, at a rate $\eta$  that just relaxes the phase build up due to the voltage, which requires
\begin{equation}
\Delta V = \pi \eta \hbar /  e \, .
\label{etaV}
\end{equation}
If, for a given current, there is a large free energy barrier, so that the rate $\eta$ given by 
Eq.~(\ref {activation}) is very small, then the resulting voltage will be proportionally small.
 
In the presence of a non-zero electric field,  there can be an additional contributions to the current from the ``normal fluid'', which arises from thermally excited quasiparticles.  When the phase slip rate $\eta$ is small, the normal current will be negligible compared to the supercurrent in a dc measurement,  but it can be an important source of dissipation in  ac applications, as discussed in Section (\ref{ac}).

Our  general considerations may be applied to a variety of situations.
In wires that are thin compared to the coherence length $\xi$, one may
neglect the variation of $\Psi$ across the diameter of the wire, and
consider that the order parameter is only a function of $x$, the
distance along the wire.  In this case, the winding number can change
if the order parameter passes through zero at some location $x_0$
along the wire.  Such events are often referred to as {\it point phase
slips}. Phase slips can also occur at  weak links, or {\it Josephson
junctions},  where the free energy has the form of a periodic function
of phase difference across the link, as opposed to the quadratic
function of the phase gradient that is most commonly applicable to a
bulk superconductor or film.

The various geometries and mechanisms for resistance by thermal activation  will be discussed below in  Section (\ref{thermal}), which  will concentrate  on  the effects of thermally activated phase slips or vortex motion on dissipation,  in single Josephson junctions, thin wires, films, and bulk materials. In Subsection (\ref{bkt}), we also discuss the role of vortices in the thermodynamic phase transition between the superconducting and normal states of a thin film. Dissipation arising from vortices induced by an applied magnetic field will be discussed in Subsection (\ref{bapp}).  

In very small Josephson junctions, 
very thin wires or highly disordered thin films,  at sufficiently low temperatures, the mechanism for
relaxation of supercurrent may be quantum tunneling of phase slips,
rather than thermal activation over a barrier.  Quantum phase slips will be discussed in 
Section (\ref{quantum}).

While the focus of this chapter is on  superconductors, the
dissipation mechanisms discussed are common to many
other types of systems. Examples can be found among neutral
superfluids such as helium and ultracold  atoms (see
Refs.\cite{packard,BECbook,dalfovo,leggett,zwerger} for reviews). For example, theoretical analyses of transport of two-dimensional systems near a superconducting or superfuid transition  
were strongly motivated by experiments on films of
$^4$He\cite{reppy}. Another interesting case, where dynamics of the
condensate order parameter determines transport properties, can be
found in bilayer quantum Hall systems at the filling factor
$\nu=1$\cite{dassarma}. In such systems one expects to find
spontaneous interlayer coherence, which is analogous to exciton
condensation\cite{eisenstein}, so that both interlayer tunneling and
antisymmetric longitudinal resistivity are determined by the
dynamics of the condensate order parameter. Current research on systems of ultracold atoms will be briefly discussed in Section (\ref{cold-atoms}).

The emphasis in this chapter will be on the theoretical concepts necessary to understand resistance in superconductors. We include only a limited discussion of experimental results, with  limited  references to the corresponding literature.  We cite more numerous references containing  detailed theoretical analyses and quantitative calculations, but here too we are far from complete.

It should emerge from the discussions below that while we believe we have a good understanding of the basic mechanisms responsible for resistance in superconductors,  there remain many puzzles and  unanswered questions about the interpretation of experimental data.  Particularly in  regimes where quantum fluctuations are important, the subject remains quite active


\section{Phase slips produced by thermal activation}
\label{thermal}

\subsection{Phase slips in Josephson junctions}
\label{thermaljj}

We begin by examining  how phase
slips arise in a Josephson junction between two finite
superconducting electrodes. 
Tunneling of 
Cooper pairs between the electrodes gives rise to the following
junction energy:
\be
U(\phi)=-E_J\cos\phi
\label{JJ1}
\ee
where $\phi$ is the phase difference between the superconudcting
electrodes, and $E_J$ is the Josephson energy (we have assumed here that any magnetic flux through the junction is negligible compared to $\Phi_s$).  
A non-zero value of $\phi$ will lead to a supercurrent across the junction, given by
\be
I_\phi = (2 e / \hbar) \partial{U}/ \partial \phi = (2 e / \hbar) E_J \sin \phi \, .
\ee
For a tunnel junction between  two BCS superconductors
with an energy gap $\Delta$, 
$E_J$ is given by the Ambegaokar-Baratoff formula:
\be
E_J=\frac{\hbar \Delta}{4e^2R_T} \tanh\l(\frac{\Delta}{2T}\rr) ,
\ee
where $R_T$ is the resistance of the junction in the normal state right above $T_c$.

If the superconducting electrodes are thick enough so that there is a large energy barrier for the order parameter to vanish inside them, then the time dependence of $\phi$ should be given by the Josephson relation 
Eq.~(\ref{josephson}), with $V$ being the instantaneous voltage difference between the electrodes.  This voltage may fluctuate in time because of thermal noise or because of quantum-mechanical fluctuations.  Here we consider only the classical thermal contributions.  

The precise dynamics of $\phi$ will depend on the way the junction is inserted in an external circuit.  Suppose that the the junction is connected to an ideal constant current source,  with  current $I$. If the phase $\phi$ at some instant  of time is such that $I_\phi \neq I$, there will be a build up of the charge difference $q$ between the two sides of the junction, given by 
$d q / dt =  I - I_{\phi}$.  This, in turn,  will give rise to a voltage difference $V=q/C$, where $C$ is the capacitance of the junction. (We neglect here the capacitance of the external circuit, which is in most cases small compared to that of the junction.) 
The voltage $V$, in turn, will give a non-zero time-derivative of $\phi$.   If the only current across the junction is supercurrent $I_{\phi}$, then we have a closed set of equations for $\phi$ and $q$ which we may write in a Hamiltonian form
\be
\frac{d \phi}{dt} = \frac{2e}{\hbar}   \frac {\partial  \cal{H}} {\partial q}\, , \,\,\,\,\, \,\,
\frac{d q}{dt} =  - \frac{2e}{\hbar}   \frac {\partial \cal{H }} {\partial \phi}\, ,
\ee
\be
{\cal{H }} = U(\phi)  - \frac{\hbar}{2e}   I  \phi  + \frac { q^2}{2 C} \, ,
\label{Heff}
 \ee
 These equations suggest that we should identify $\hbar q/2e$ as a momentum canonically conjugate to $\phi$, which is consistent with the commutation  relation 
 $[\phi, q] = 2ie$ that one would expect based on Eq.~(\ref{commrel}).

\begin{center}
\begin{figure}
\includegraphics[width=10cm]{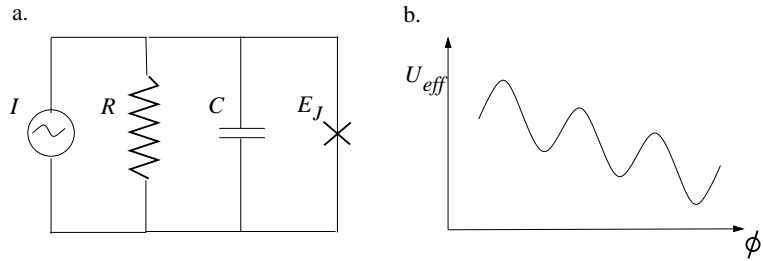}
\caption{Josephson junction connected to a current source with current
  $I$.  Panel (b) shows the  RSJ model, where the tunnel
  junction is shunted by a resistance $R$ and a capacitance $C$.  In
  most cases,  $C$ is determined by the internal capacitance of the
  junction, while $R^{-1}$ may be the sum of an external shunt
  conductance and an internal  conductance due to tunneling of normal
  quasiparticles at non-zero temperatures. Panel (b) shows the ``washboard" effective potential,  $U_{\rm{eff}}(\phi) = - E_J \cos \phi - (\hbar / 2e) I \phi$, for the phase $\phi$.
\label{WB}}
\end{figure}
\end{center}

 If $I$ is smaller than a critical current $I_c= 2 e E_J / \hbar$, the $\phi$-dependent terms in 
 $H_{\rm {eff} }$ have the form of a ``washboard potential", illustrated in Fig.~(\ref{WB}a), with a local minimum
 and a local maximum in each unit cell. The local minima are stable solutions of the equation $I_{\phi} = I$.  If the system is initially placed at such a point, and it initially has $q=0$, then in the absence of external noise it will stay there forever, with a constant current $I$ and $V=0$.  If the system is displaced slightly from one of the minima, it 
 will oscillate forever about this minimum,  unless some additional coupling supplies energy enough to get it the over barrier separating it from one of the neighboring minima.  As long as the system is trapped near one of the minima, the time-average of $d \phi/dt $ will still be zero, so we again have a current carrying junction with $V=0$.

When $I=0$, the energy  necessary to go over a barrier  will be equal to $ 2 E_J$, regardless of whether the system moves in the direction of increasing or decreasing $\phi$. When $I \neq 0$, however, the barriers $\Delta F^{\pm}$ for increase or decrease of $\phi$ will differ from each other by an amount $- \pi \hbar I / e$. If $I/I_c \ll1$, one finds simply
 \be
 \Delta F^{\pm} \approx  2 E_J \mp \frac {\pi \hbar I }{2 e} \, .
 \label{fpm}
 \ee
 
 The energy to get over the barrier can come from coupling to thermal fluctuations in the external circuit, or from thermally excited quasiparticles tunneling across the junction. 
 At non-zero temperatures, in addition to the supercurrent, there will generally be a normal current across the junction, due to thermally excited quasiparticles, when the voltage difference $V(t)$ is different from zero. This can be represented, at least approximately, by a model in which there is a shunt resistor, with resistance $R$, across the Josephson junction, as shown in Fig. (\ref{WB}b).    If the external circuit is not an ideal current source, then $R^{-1}$ should be a sum of the contribution from the quasiparticles and the differential conductance of the external circuit.   The shunt resistance will lead to  an added term, 
$-q/RC$ in the equation for $d q/dt$, which will lead to damping of the coupled oscillations  $q$ and $\phi$.  In addition, one must include a white noise term in the equation of motion, proportional to $T/RC$, so that for $I=0$, the system can reach thermal equilibrium, with a probability distribution 
$P(\phi) \propto \exp [-U (\phi) / T]$.   If one uses the same white noise  source for $I \neq 0$, one finds a transition rate, in the direction of  increasing or decreasing $\phi$, which may be written in the form
\be
\eta_{\pm} =  \Omega e ^ {- \Delta F^{\pm} / T} \, ,
\label{etapm}
\ee
 where the prefactor $\Omega$ depends, in general on the parameters $R$ and $C$, as well as on the Josephson energy $E_J$.  In the limit $I/I_c \ll1$ one can neglect the difference in the prefactors for forward and backwards transitions.  Then the mean voltage across the junction,   which  is proportional to the net difference $\eta = \eta_+ - \eta_-$, in accord with Eq. (\ref{etaV})   is given by
 \be
 \sex {V} =   ( 2\pi \hbar /e ) \,  \Omega \, e ^{-2E_J / T} \sinh  (\pi  \hbar I / 2 e T)  \, .
 \label{vjj}
 \ee
 It the limit of small currents $(I \ll e T / \hbar)$ one obtains an effective dc resistance for the junction:
 \be
 R_{\rm{eff}} =  \sex{V} / I =   (\pi \hbar^2  /  e^2 T)  \, \Omega e ^{ -2E_J / T} 
 \label{reff}
\ee 

When the barrier is large compared to $T$, so that $R_{\rm{eff}}$ is small compared to the shunt resistance or the normal resistance of the junction, then essentially all of the current through the junction is  the supercurrent, and the total current $I$ is essentially the same as the mean value of  $I_\phi$.  

It should be noted that the barrier $\Delta F^+$ tends to zero as $I \to I_c$.  Thus, even if $E_J / T$ is quite large, so that phase slips are negligible for $I / I_c \ll 1$, there may be a regime close to $I_c$ where thermally activated phase slips become important.

The prefactor $\Omega$ can actually be calculated for the RSJ model. In the absence of damping, a junction near a minimum of the cosine potential  $U(\phi)$ will oscillate about it at  the Josephson plasma frequency
\be
\Omega_{JC} \equiv  \frac{1}{\hbar}  \left( \frac{2 e^2 E_J}{C} \right)^{1/2} .
\label{ojc}
\ee
Then for an underdamped junction, where $ \Omega_{JC} RC$ is larger than 1, (but not larger than 
$E_J/T$), in the limit $I \to 0$, the pre-factor, may be written\cite{IvanchenkoZilberman,AmbegaokarHalperin}
$\Omega =  \Omega_{JC} / 2 \pi$.
For an overdamped junction, with $ \Omega_{JC} RC \ll 1$, one finds in the limit 
 $I \ll I_c$, that\cite{AmbegaokarHalperin} 
 \be
 \Omega = \frac {e^2 E_J RC}{ \pi \hbar^2 } .
 \ee

\subsection{Thermal phase slips in a thin wire close to $T_C$.}
\label{thermalwires}

When  a superconducting wire is narrow compared to the coherence length $\xi (T)$, it is generally correct to neglect variations in the order parameter across the diameter of the wire, and to treat $\Psi$, at any time $t$, as as a function of a single position variable $x$, the distance along the wire. A vortex, in this case, degenerates to a single point where $\Psi(x)=0$, and a phase slip event is an isolated point in space and time, where $\Psi$ passes through zero and the phase difference along the wire jumps by $\pm 2 \pi$.  

For temperatures relatively close to $T_c$, one may use the Ginzburg-Landau functional   (\ref{GL}) to calculate the free energy barriers $\Delta F^{\pm}$ for creation of a phase slip in the wire.  In analogy to our treatment of the Josephson junction, if the wire is connected to an external current source with a current $I$, we must add to the GL functional a term $- (\hbar/2e) I \Delta \phi$, where $\Delta \phi$ is the difference in the phases $\phi(x)$ measured at the two ends of the wire.  (We may assume that $\Psi \neq 0$ at the two ends, so that the phases at the ends may be defined to be continuous functions of time.)  Alternatively, one may calculate the free energy barrier to produce a phase slip in a closed superconducting loop that carries a current $I$.

The free energy  activation barrier for phase slips in a wire was first worked out, using the GL theory,  Langer and
Ambegaokar \cite{LA}. 
For a wire loop of length L , with a specified phase change $2 \pi n $, 
 the  order parameter configuration that  leads to a local minimum of the free energy functional has a uniform phase gradient $k = 2 \pi n / L$ and a constant magnitude of $\Psi$, with 
\be
\Psi_k(x)=\sqrt{\alpha/\beta}e^{ikx}(1-\xi^2 k^2)^{0.5}. 
\ee
A phase slip will reduce $k$ by $2\pi/L$. 
For that to occur, however, the order parameter must go through an intermediate
state $\Psi_s(x)$ which is a saddle point of the G-L free energy,  where at some point $x_0$ in  the wire, the order parameter nearly
vanishes. The  unwinding of the phase is concentrated close to  $x_0$, and the magnitude of the order parameter is depressed in a region of lengnth $\xi$ about that point.
point.    An explicit solution for  $\Psi_s(x)$, for arbitrary current,   was given by McCumber and Halperin.\cite{MH} 
Once the order-parameter configuration reaches the saddle-point configuration $\Psi_s$, the configuration can evolve with a continuously decreasing free energy  through the actual phase slip event,  where the order parameter passes though zero at some point on the line. 

 Roughly speaking, the free
energy cost of reaching the saddle point is the cost of
suppressing the order parameter in the phase slip region. More precisely, in the limit of zero current, the barrier is found to be
\be
\Delta F_0 = K_0 A \xi f_0
\label{DFMH}
\ee
where $A$ is the cross-sectional area of the wire,   $K_0 = 8 \sqrt{2}/3 $, and $f_0=\alpha^2/ 4 \beta$ is the condensation energy per unit volume.  (The quantity $f_0$ is related to the critical field $H_c$  by
$f_0 = H_c^2/8 \pi$.) The barrier $\Delta F_0$ decreases $\propto (T_c - T)^{3/2}$ when $T \to T_c$.

For currents that are non-zero but small compared to the critical current, {\it i.e.}, $k \ll \xi^{-1}$,  the barriers for forward or backwards phase slips have the form 
 $\Delta F^{\pm} = \Delta F_0 \mp \pi \hbar I / 2e$, as was found for the single Josephson junction. 
Then, if we assume that the rate of phase slips has an activated form similar to Eq. (\ref{etapm}), we obtain a  voltage drop $\sex{V}$ identical to (\ref{vjj}), with the zero-current activation energy $2E_J$ replaced by $\Delta F_0$.

In order to calculate the pre-exponential factor $\Omega(T)$,  one must make additional assumptions about the dynamic equations of motion for the order parameter.  
This was done by  McCumber and Halperin \cite{MH} using the simplest possible model, the
time-dependent Ginzburg-Landau (TDGL) theory.  In this model, the order parameter obeys an equation of motion of the form
\be
\frac{1}{\Gamma}   \der{\Psi}{t}=-\frac{\delta
  F}{\delta\Psi^*}
    +\eta(t)
\label{tdgl}
\ee
where $\eta(x,t)$ is a Gaussian white noise source with
 correlator $\langle
\eta(x,t)\eta(x',t')\rangle= 2k_B T \, \Gamma^{-1} \delta(t-t') \delta(x-x')  $, and 
\be
\frac{\delta
  F}{\delta\Psi^*}=\frac{\gamma}{2}\l(\nabla+\frac{2ie}{\hbar
  c}A\rr)^2\Psi-\frac{\alpha}{2}\Psi+\frac{\beta}{2}\Psi \, |\Psi|^2 \, .
\ee
The constant $\Gamma$ is  temperature-independent near $T_c$, and is chosen  so that 
$\Gamma \alpha = 8 k_B (T_c-T) / \pi \hbar \equiv 1/ \tau_{GL}$.  With these equations, one finds that in the absence of a current, the function $\Psi(x)$ will have an equilibrium distribution 
$P \{\Psi \} \propto e^{-F/T}$, and that an initial state close to equilibrium will relax to it at a rate $1/\tau_{GL}$, which goes to zero as $T \to T_c$.

The explicit calculation of the activation rate from  the TDGL equation uses  not only of the values of $F$ at the minimum and the  saddle point, but also the eigenvalues of the second derivative matrices at the two points.  The positive eigenvalues contribute  entropy corrections,  which modify the numerical value of $\Omega$, while the negative eigenvalue at the saddle point determines the overall time scale.  
For a translationally invariant system, such a calculation 
always
results, up to a factor of order unity, in the product of the number of ``independent''
phase-slip configurations (e.g., where along the wire phase slips can occur), the inverse of the
TDGL time constant, and the square root
of the free-energy cost of the saddle configuration divided by the
temperature \cite{GolubevZaikin}. In the limit $I \to 0$, the explicit result for the prefactor, obtained by McCumber and Halperin, is
\be
\Omega(T) =\frac{ 0.156 } {\tau_{GL}} \frac {L}{\xi} \left( \frac{\Delta F_0}{T} \right)^{1/2} \, ,
\label{preMH}
\ee
where $L$ is the length of the wire.  
For $T \to T_c$, the prefactor varies as $(T_c-T)^{9/4}$. 
The electrical resistance is resistance is given by 
$R=(\pi \hbar^2/e^2 T) \Omega e^{- \Delta F_0/T}$,   in analogy to 
(\ref{reff}), using  (\ref{DFMH}) and (\ref{preMH}).

 Above $T_c,$ the  order parameter fluctuations predicted by TDGL theory give a temperature-dependent contribution to the electrical conductivity, in one, two and three dimensions, which was first calculated by Aslamazov and Larkin.  An interpolation between this regime and the McCumber-Halperin formula therefore gives a prediction for the entire temperature dependence near $T_c$, within the TDGL theory.

The McCumber-Halperin formula for the resistance of a wire seems to work
surprisingly  well in fitting resistivity data on thin wires close to $T_c$  (see, e.g., \cite{Newbower}), 
and even at lower temperatures\cite{Bezryadin_LAMH} .  Nevertheless, the
formula comes with some caveats. 
The TDGL theory is
derived by integrating out the fermionic degrees of freedom, and
assuming that they can respond rapidly to changes in the 
order parameter. This is probably a good assumption for calculating time-dependent fluctuations in the order-parameter
 at temperatures slightly above $T_c$.  However, its use below $T_c$ is  problematic.  One problem is that the energy gap of most superconductors becomes  rapidly bigger than $k_B T$, at a small distance below $T_c$.  Even where this has not occurred, there  can be some very long relaxation times associated with small rates for inelastic scattering of quasiparticles, restoration of ``branch imbalance'', etc.  In principle, these effects could lead to a very large reduction in the value of  the prefactor, which could change considerably the interpretation of the data.  Even above $T_c$ , there are  additional contributions to the electrical conductivity in addition to the  Aslamazov-Larkin term which may be important in the experiments.  The overall situation remains to be sorted out.

\subsection {Planar geometries.}

For a thin film superconductor, whose thickness $d$ is small compared to the penetration depth $\lambda$, we can generally neglect the effects of magnetic fields produced by currents in the film.  Therefore we can take $\vA$ to be the vector potential due only to an applied external magnetic field.  In the absence of such a field, we may drop $\vA$ from the equations. Moreover, if $d$ is small compared to $\xi$, we may neglect variations in $\Psi$ over the thickness of the film.   One may then look for the planar configuration $\Psi(x,y)$ that minimizes  (\ref{GL})  subject to the constraint that there exist  a pair of vortices of opposite sign, with a given separation $2\vD$, which we choose to be large compared to the coherence length $\xi$ but small compared to the distance to the nearest boundary.  The resulting free energy,  relative to the free energy of the ground state without  the vortices, is  
\begin{equation}
\delta F = 2 \pi K d [ \log (D/\xi) + 2 \epsilon_c]  \, , \, \, \, \,
K =  \gamma \alpha / \beta \, , 
\label{pair}
\end{equation}
where $\epsilon_c$ is a constant of order unity.
If there is a non-zero background supercurrent density $\vj$, however, resulting from a uniform gradient of the phase $\phi$, one finds an additional term in the free energy of the form  
\begin{equation}
\delta F_j = - (\pi \hbar /e)  j  D_{\perp} d  \, ,
\label{jterm}
\end{equation}
where $D_{\perp}$ is the component of the separation perpendicular to
the current. This additional term may be derived by adding a term
$-(\hbar/2e)I \Delta \phi$ to the effective Hamiltonian, in analogy to
the our treatment of the Josephson junction and one dimensional wire
where $\Delta  \phi$ is the phase difference between the two ends of
the film and $I$ is the total current, integrated across the width of
the sample. Minimization of the free energy in the presence of a pair of vortices leads to an incremental change in $\Delta \phi$ which is equal to $2 \pi D_{\perp} / W$, where $W$ is the width of the film.  The additional term may also be 
understood as arising from the magnus force on a vortex, , which is proportional to the circulation of the vortex and the background current, and is perpendicular to both.
Similar formulas apply to a single vortex interacting with its own image charge near an edge of the superconductor, except that in this case the right hand sides of 
(\ref{pair}) and (\ref{jterm}) must be divided by two, while $\vD$ is the distance to the edge.  In either case, we see that if the displacement $\vD$ is increased in the direction of the magnus force, the free energy will eventually decrease, as the linear term will win out over the logarithm.   The free energy maximum occurs when $D_{\perp} \approx 2 e K /  j \hbar .$
The free energy barrier for nucleating a vortex at one edge of a sample, and freeing it from its image charge, is thus found to be, for small $j_s$, 
\begin{equation}
\Delta F  \approx \pi K d \log (2e K  / j \hbar  \xi) .
\label{feb}
\end{equation}
At low temperatures, $T\ll K d$, this results in a dissipative
response which gives a voltage of the form
\be
V \propto e^{-\Delta F/T}  \propto  j^{x(T)}.
\label{RT1}
\ee
where $x(T) \approx \pi K d / T$.  A more accurate analysis, which includes the contribution to the pre-exponential  factor arising from the entropy associated with  the position of the vortex, predicts that
\cite{HalperinNelson, AHNS} 
\be
x(T) \approx 1 + \pi K d / T 
\label {RTexp}
\ee

Thus vortex nucleation processes produce dissipation in a superconducting strip at any finite
temperature and current. However, if $x(T)$ is large, the differential resistance arising from
Eq. (\ref{RT1}) vanishes as a large power law as the supercurrent approches zero. We shall see that the exponent is always larger than 3 in the superconducting phase.

In a film of finite width, the logarithmic increase of $\Delta F$ will
be cut off when the current is so small that  $D_{\perp} $ reaches
$W/2$.  Also, if one takes into account the magnetic field produced by
a vortex, the increase in $\Delta F$ if  $D_{\perp}$ exceeds  the
length scale $\lambda_{\perp} = 2\lambda^2/ d$ for magnetic screening
in a film (also known as the Pearl penetration length). Thus for sufficiently small currents, one should recover a linear voltage-currrent relation, with a small value of the resistivity.

\subsection{Thermally excited vortices and the BKT transition}
\label{bkt}

At temperatures comparable to the phase stiffness  $ K d$      of a superconducting film,
vortices may arise as thermal excitations. In an
infinitely thin film, these vortices are described as an interacting gas of
Coulomb particles, i.e., with an interaction that depends
logarithmically on distance (in a film with a finite thickness the
logarithmic interaction will prevail at distances shorter than the
Pearl penetration length of $\lambda_{\perp}=2\lambda^2/d$. The bare fugacity of the vortices is dictated by their core
energy, $E_c=2\pi K d \epsilon_c$, and is given by:
\be
\zeta\approx \frac{1}{\xi^2} e^{-2\pi K d \epsilon_c/T},
\ee
which may be taken as a rough estimate for the total density of vortices in a
film. 

The intervortex interaction leads to a subtle vortex-pairing
transition, known as the Berezinskii-Kosterlitz-Thouless transition \cite{KosterlitzThouless,Berezinskii,Nelson}. At
high temperatures, the vortices behave as a charged, but unbound,
plasma. As the temperature drops under a critical value, $T_{KT}$,
vortices form neutral vortex and anti-vortex bound pairs. 
The temperature where this transition occurs can be obtained by simple thermodynamic
consideration\cite{Kosterlitz}. If we consider a finite, but large, system of size $L$,
we can compare the energy cost of adding an uncompensated
vortex with its entropy gain. The energy increase due to an
uncompensated vortex in a film of side $L$ follows from
Eq. (\ref{pair}), and is given by $\Delta U=\pi K d [ \log (L/\xi) + 2
  \epsilon_c]$. On the other hand, such a vortex has an entropy which
is roughly $\Delta S=\ln (L^2/\xi^2)=2\ln L/\xi$. The net increase of
free energy through the nucleation of a vortex is thus:
\be
\Delta F\approx (\pi K d - 2T)\ln L/\xi
\label{dfv}
\ee
where we ignore the film-size independent core energy. We see that this free energy cost diverges in the thermodynamic limit, if $T< T_{KT}$,  defined by
\be
T_{KT}=\frac{\pi K d}{2}.
\label{TKT}
\ee
Vortices are  then logarithmically confined into neutral pairs, and free vortices do not exist.  The system is still a superfluid, though the value of $\rho_s$, and hence of $K$ will be somewhat reduced because of the polarizability of bound vortex pairs in the presence of a current.  [It is this temperature-dependent renormalized value of $K$  that  should be used in   (\ref{TKT}) to determine $T_{KT}$.]
For $T>T_{KT}$,  single vortices would proliferate and form a
free plasma.  As free vortices should have a finite diffusion constant, they would move in response to an arbitrarily small electrical current, giving a  non-zero resistivity.

Another perspective on the BKT transition is obtained by considering
the spatial correlations of the superconducting order parameter in
thin films. 
Because   two dimensions is the lower critical
dimension for the $U(1)$ superconducting order parameter, there is not true long range order at non-zero temperatures.  Instead, the correlation function in the superfluid phase is predicted to decay slowly at large distances, according to the power law, 
\be
\langle e^{i(\phi(r)-\phi(0))}\rangle\sim \frac{1}{r^{T/\pi K d}}  .
\ee
(The exponent here is necessarily $< 1/4$ in the superfluid phase.)  For $T>T_{KT}$ the correlation function falls off exponentially with a decay length $\xi_+(T)$.

A proper analysis of the BKT transition makes use of a renormalization group (RG) approach, in which one repeatedly integrates out the effects of pairs separated by a distance smaller than a  running cutoff 
of form $a_l = \xi e^l$, and keeps track of the renormalization of $K$ and the vortex fugacity $\zeta$.  For $T>T_{KT}$ the renormalization must be  stopped at a length scale where the density of vortices is comparable to $a_l^{-2}$, and the correlation length $\xi_+$ is identified with this value of $a_l$.  The RG analysis predicts specifically that $\xi_+$ should diverge at $T_{KT}$ with an essential singularity:
\be
\xi_+ \sim e^{-a/\sqrt{|T-T_{KT}|}},
\label{xiplus}
\ee
where $a$ is a constant.
The confinement length of the bound vortex pairs should diverge in a similar manner for $T \to T^-_{KT}$.
Because the correlation length diverges so rapidly above $T_{KT}$, there should only a very weak essential singularity in the specific heat at the BKT transition point itself: all derivatives of the free energy should be continuous there.   

The linear resistivity for $ T > T_{KT}$, should be  proportional to  the density of free vortices $\sim 1/ \xi_+ ^2$.  
For  $T < T_{KT}$, the  resistivity vanishes  in the the limit $j \to 0$, but finite currents can cause disassociation of bound vortex pairs, producing a non-zero voltage.    For small currents, this voltage should be described 
by the power law (\ref{RT1}). Comparing (\ref{RTexp}) with (\ref{TKT}), we see that   the exponent $x(T)$ is
$>3$ for $T<T_{KT}$, and it approaches 3 for $T\to T_{KT}^-$.  One also predicts
\cite{HalperinNelson,AHNS}
 that for $T=T_{KT}$, the induced  voltage should have  the universal dependence  $V \propto j^3$.

The BKT transition as presented above applies equally well to these films of superfluid $^4$He as to thin-film superconductors. Indeed the theory has been supported  by a number of beautiful experiments in the helium case.\cite{reppy} 
An important feature of the superconductor, however, is that for a bulk sample, the thermal variation of the superfluid density and other parameters is very well given by a mean field theory, such as the BCS or Ginzburg-Landau theories, except in an extremely narrow range near the bulk transition temperature $T_{c0}$.  The transition temperature $T_{KT}$ in a film is far enough below $T_{c0}$ that the mean field theory may be used to estimate the bare parameters of the RG analysis. The parameter $a$ in (\ref{xiplus}), as well as the pre-exponential factors, can be estimated from the mean field parameters and the measured normal conductivity $\sigma_n$. \cite{HalperinNelson} Similarly, using the time-dependent Ginzburg-Landau theory, one may estimate quantitatively the enhancement of the  conductivity above $\sigma_n$ due to incipient superconducting fluctuations for a range of temperatures above $T_{KT}$.  

A number of experiments on
thin superconducting films have indeed observed the predicted forms of the resistivity 
\cite{Bezryadinfilms, FioryHebard}. However, there are also experiments, particularly involving cuprate superconductors, which have been fit to different functional forms. A detailed analysis by Strachan. Lobb, and Newrock
\cite{SLN} suggests that the apparent discrepancies may be due to a combination of uncertainties in the choice of the mean-field $T_c$ and problems where the vortex separation length  $D_\perp$ may be exceeding $\lambda_\perp$ or the film width.  The reader is referred to that article for a detailed discussion, as well as for citations to the experimental and theoretical literature.


\subsection{dc Resistance in bulk superconductors without magnetic fields}

Bulk superconductors support a more robust superconducting
state, and dissipative effects in them are much more suppressed than in
lower-dimensional superconductors. We first consider a wire that is thick compared to the coherence length $\xi$, but thin compared to the  magnetic penetration depth $\lambda$, which is possible in an extreme type II material.  Then the  mechanism for producing  dissipation  the presence of a supercurrent  density $j$ is the thermal nucleation of vortex rings, which can expand across the diameter of the wire, and change the phase by $2 \pi$.  When the vortex ring is small, the line tension due to the vortex energy will try to collapse the ring, but eventually, for a large vortex, this will be overcome by the  Magnus force due to the current, which will  favor expansion of the ring.

The rate of vortex loop formation is thus inhibited by an energy barrier,
which thermal fluctuation may overcome
\cite{FisherFisherHuse}. Following the same logic that led to
Eq. (\ref{RT1}), we note that the energy of a vortex ring of radius
$R$ is:
\be
E_{ring}  \approx 2\pi R E_c - \pi R^2 j \Phi_0,
\label{vringE}
\ee
with $E_c$ being the vortex energy per unit length. Note that $E_c$ also
contains a slowly varying contribution $\sim \ln(R/\xi)$ when
$\xi<R<\lambda$; this does not affect the scaling behavior we discuss.
The vortex-ring  energy has a maximum at
$R_{max}=E_c / j\Phi_0$ with energy $E_{B}\approx \pi E_c^2 /   j\Phi_0$.
From this energy barrier we infer that the finite-current resistivity of a
bulk superconductor should vary as 
$
\rho \sim e^{-J_T/j}
$
with:
\be
J_T=\frac{\pi E_c^2}{T\Phi_0}.
\ee
This  leads to a rapidly vanishing linear resistivity in
the limit of $j \rightarrow 0$ as expected. In practice, this exponential behavior is so strong, that there would typically be a threshold, say $J_T / 30$, below which no voltage drop could be measured in practice.  

At sufficiently high current densities, when
$j $ exceeds the Ginzburg-Landua critical current $J_F\sim \frac{E_c}{\Phi_0\xi}$, superconductivity would
break down not due to thermal fluctuations, but due to the mean-field
energy cost of the current, according to the G-L free-energy
functional, Eq. (\ref{GL}).  ($J_F$ is also  the current density  where the
energy-maximum radius would be comparable to the coherence length, $\xi$,
so vortex rings stop being a useful concept).   
The condition $J_T<J_F$,  occurs only extremely close to $T_c$, in the Ginzburg
fluctuation regime where mean field theory breaks down, and  even  the  condition  $J_T/30 <J_F$  for measurable dissipation due to formation of vortex rings may  be difficult to achieve.

For a superconducting wire that is thicker than the magnetic penetration depth, the situation is more complicated because the current is confined to a  skin depth $\lambda.$  This can  make it even more difficult for a vortex ring to expand across the interior of the wire, where the current density is small. Again, if the current is sufficiently low, one can easily enter the regime where  dissipation rates are unmeasurably small, in the absence of applied magnetic fields.

\subsection {Resistance in an applied magnetic field}
\label{bapp}


For a bulk type II superconductor, when the applied magnetic field exceeds the lower critical  field $H_{c1}$ of the material, the magnetic field will penetrate the superconductor, giving rise to a finite density of vortices, even in thermal equilibrium.  If there is also a non-zero macroscopic electric current flowing in the superconductor, the vortices will feel a magnus force, which tries to move them in the direction perpendicular to the current flow.  If the vortices can  move in response to this force, there will be an induced voltage, proportional to the density of vortices and their mean velocity of motion, which will lead to dissipative resistance.

If the current is not too large, and if the temperature is low, the net rate of vortex motion can be extremely small, as the vortices will tend to be pinned by imperfections in the material.
The origin of pinning can be point-like crystal defects due to interstitial atoms, impurities or
vacancies, or due to extended defects such as grain boundaries, twin boundaries, or dislocations. Variations in material composition or crystal structure could also cause pinning.   In a thin film, variations in the film thickness could lead to variations in  the core energy which could  lead to 
pinning. For applications of superconductors as high current
transmission lines or in superconducting magnets, one wants to have
pinning forces that are as strong as possible, to prevent the motion
of vortex lines and to suppress dissipation. 
For this reason, pinning sites are often added by adding impurities or
by cold working. 
Pinning is most effective when the pinning objects have size comparable to the coherence length $\xi$, which is the size of the vortex core.


In the absence of disorder, vortices in a bulk superconductor in a magnetic field will tend to form a triangular array, known as an Abrikosov
vortex lattice \cite{abrikosov}.    Since the interactions between  vortices try to keep them in a regular array, vortex motion in the presence of a macroscopic current  is
actually the result of the competition between the magnus force, the lattice stiffness,
and the random pinning potential. This interplay was qualitatively discussed
by Larkin and Ovchinnikov \cite{LarkinOvchinnikov} [LO] under the assumption that the pinning is caused by a large density $n_p$ of pinning objects, each having a weak effect on the vortex lattice. The energy gain for  a pinning site inside a vortex core was assumed to have a value $u$ for a site at the center of the core, falling to zero smoothly over the core radius.  Then the root-mean square pinning force  exerted by a pinning at a  random position is given by $f_p \approx u/a$, where 
$a \approx  (\Phi_0/B)^{1/2}$ is the distance between vortices in the Abrikosov lattice.  LO assume that 
the lattice is fragmented into domains where the lattice order is maintained, while
the lattice distorts slightly to accomodate the random pinning
potential, and they estimate the domain size and shape that will
optimize the free energy gain due to the pinning potentials.  They
find that the optimum domain volume $V_c$ and the pinning energy $\delta F_p$ for each domain are given by \cite{TinkhamBook}:
\be
V_c \approx \frac{ C^2_{tilt} C^4_{shear} \xi^6}{n_p^3 f_p^6}\, ,  \, \, \, \, \, \, 
\delta F_p \approx   - \frac{ n_p^2 f_p^4}{C_{tilt} C^2_{shear} \xi^2}
\ee
where  $C_{tilt}=BH/4\pi$ is the tilt modulus of the Abrikosov lattice, and
$C_{shear}\approx (1-B/H_{c2})^2 B H_{c1}/16\pi$ is its shear modulus.  LO assume that the maximum pinning force per unit volume  $\approx f_p (n_p/V_c)^{1/2}$ determines the critical current $J_c$ for the onset of vortex motion at zero temperature, which leads to the result
\be
J_c B \sim \frac{n_p^2
f_p^4}{C_{tilt}C_{shear}^2\xi^3}.
\ee
Following the same arguments for a superconducting film yields:
\be
J_c B \sim \frac{n_p
f_p^2}{C_{shear}\xi d}.
\ee

It is interesting to note that according to this logic any amount of
pinning would  render a superconductor dissipationless for sufficiently low currents. However, in the LO weak pinning limit, the stiffer the vortex lattice,  the smaller is the critical current.  This is different from the case of a small density of very strong pinning sites, where the lattice stiffness could enhance the effects of pinning.

At finite temperature vortex motion may arise below $J_c$ due to thermal
fluctuations.  Even a
qualitative understanding of these effects is difficult since it
requires understanding not only the characteristic forces that
vortices encounter, but the shape of the collective pinning potential
as a function of current. Early collective flux pinning theories
proposed by Anderson and Kim \cite{AndersonKim} suggested that the
maximum potential barrier for a vortex to depin is $U(j)\propto
(1-j/J_c)$. A more modern approach \cite{FisherFisherHuse} emphasizes
the fact that the potential barrier, due to collective effects,
actually diverges near $J\rightarrow 0$ as:
\be
U(j)\sim U_0\l(\frac{J_c}{j}\rr)^{\mu},
\label{muJ}
\ee
where $\mu$ is an exponent $\leq 1$. 
The voltage drop $V$ in a superconductor with current density $j$ should then be  proportional to the Boltzmann factor, giving 
\be
V(j)\sim \exp \left(-\frac{U_0}{T}\l(\frac{J_c}{j}\rr)^{\mu}\right).
\ee
This modified Arrhenius law leads to a vanishing linear resistance in
the limit of zero current, but any finite current at a finite
temperature will experience  dissipation.  Combining this with the   Anderson-Kim model near $j\sim J_c$, Eq. (\ref{muJ}), leads to the prediction that a  superconducting loop that initially carries a current close to $J_c$ will experience a current decay which becomes slower and slower with time, 
with a form 
\be
j(t)\approx J_c\l(1+\frac{\mu k_B T}{U_0}\ln (1+t/t_0)\rr)^{-1/\mu},
\ee
where $t_0$ is a microscopic time scale.
 
If disorder is sufficiently strong, it may obliterate the Abrikosov
lattice altogether. In this case it is thought that at low
temperatures a {\it vortex glass} phase replaces the Abrikosov lattice
phase \cite{FisherFisherHuse}. The properties of this phase, which has
many concequences for high $T_c$ cuprate superconductors, lie beyond
the scope of this review. 
An exhaustive review article on the rich topic of
vortex pinning and motion has been given by  Blatter et al. \cite{LarkinReview}.  

For currents larger than $J_c$ at low temperatures, and even for weak currents close to $T_c$ if the disorder is small, the vortex lattice may be unpinned from the defects, and may flow freely under the Magnus force of the current.  We would then expect to find a vortex drift velocity $v_d$ proportional to the current density $j$, with a coefficient $\eta^{-1}_v$ that depends on the temperature and the material at hand, but may be relatively insensitive to the quantity of defects. (We are concerned here only with the component of motion parallel to the Magnus force, which means perpendicular to $\vj$.) 
This gives an electrical resistivity, in the flux flow regime, given by
\be
\rho = \frac {\pi \hbar B}{\eta_v e \Phi_0} = \frac {B}{\eta_v c} .
\ee

A crude estimate of $\eta_v$ was obtained by Bardeen and Stephen, who modeled the vortex core as a region of normal fluid, and estimated the rate of energy dissipation in the core of a moving vortex by calculating the normal current induced by an effective field proportional to  product of the drift velocity and an effective magnetic field of $\Phi_0/ \pi \xi^2$.   This led to an estimate 
\be
\eta_v \approx \sigma_n \Phi_0 / \pi \xi^2 c,
\ee
where $\sigma_n$ is the electrical conductivity of the normal metal.  Using the relation between $\xi$ and the upper critical field    $H_{c2}$ and $\xi$, we find an approximate form for the flux-low resistivity:
\be
\rho \approx \frac {B} { \sigma_n H_{c2}} .
\ee
Despite the crude approximations involved, this formula seems to work surprisingly well in many cases.



\section{Quantum fluctuations in junctions, wires, and films}
\label{quantum}

In mesoscopic superconducting devices, phase slips may occur due to
quantum flucutations rather than thermal fluctuations. As mentioned above, the phase of the superconducting order parameter is canonically
conjugate to the charge density, or Cooper pair density. Therefore
charging terms in the Hamiltonians describing such superconducting
devices will produce fluctuations of the phase variable, and lead to
dissipation. In superconducting wires and higher dimensional arrays
the competition between charging effects and the Josephson coupling
terms may give rise to a zero-temperature quantum phase transition,
and not just to finite current dissipation. In this section we will
touch upon these effects. 

\subsection{Quantum phase slips in Josephson junctions}

Let us return to the model of a capacitative Josephson junction connected to an ideal constant current source, considered in Section (\ref{thermaljj}).  For the moment we assume there is no shunt resistor or other sources of dissipation, so that the system may described by the Hamiltonian $\cal{H}$ defined in (\ref{Heff}   ), with the previously stated commutation rule
$[\phi, q] =2ie$.  We shall now treat the problem quantum mechanically, however, rather than taking  the classical limit.  It is convenient to think about wave functions which depend on the variable $\phi$, so that $q$ is represented by the operator $q= -2e i \partial / \partial \phi$.

In a capacitatively shunted Josephson junction, quantum mechanics has
several effects.  We shall focus on the case were 
$E_J > E_C$ where  $E_C=2e^2/C$ is the Cooper pair Coloumb blockade energy, and we shall first consider the situation where the external current $I=0$.  Suppose that the system is initially 
trapped near  the cosine minimum at $\phi=0$, so that
 $ -E_J\cos\phi\approx\frac{1}{2}E_J\phi^2-E_J$.   
 The
approximate Hamiltonian is now that of a harmonic oscillator, with 
resonance frequency given by Eq. (\ref{ojc}), 
$\Omega_{JC}= (E_J E_C)^{1/2} /\hbar$. Quantum mechanics predicts a series of energy levels, separated by $\hbar \Omega_{JC}$, near the bottom of each  cosine well.

The second quantum effect is the possibility for $\phi$ to tunnel  between two adjacent wells.  This process is a quantum phase slip, and
its amplitude can be estimated from a WKB calculation:
\be
\zeta\sim  \Omega_{JC}  \sqrt{S}  e^{-S}
\label{zetadef1}
\ee
where the action barrier $S$ is given by  \cite{GolubevZaikin} 
\begin{eqnarray}
S=-\intt_0^{2\pi} d\phi \left( \frac { E_J (1-\cos \phi)}{E_C} \right)^{1/2} 
 =4\sqrt{2}\sqrt{E_J/E_C}.
 \label{QPSaction}
\end{eqnarray}

For a Josephson junction consisting of two  superconductors separated by an insulating layer of fixed thickness,  the coupling energy $E_J$ will be proportional to the area $A$ of the junction, while $E_C \propto A^{-1}$.  Thus the frequency $\Omega_{JC}$ is independent of $A$, while the action $S$ is proportional to $A$.  We shall be interested in small junctions,  where $S$ is larger than unity but not so large that the tunneling rate is completely negligible.

We can now  ask at which temperatures does the quantum tunneling
process become more pronounced than the thermal phase slip, whose rate is given by (\ref{fpm}) and (\ref{etapm}).
 We see that quantum tunneling should become more important than thermal slip processes when 
$e^{-2E_J/T}\ < e^{-S}$, which means that  $T$ should be less than the crossover temperature $T_Q = \hbar \Omega_{JC}$.  

To treat the tunneling more quantitatively, we may use the analogy with  a particle in a periodic potential. When $I=0$, the eigenstates of the  Hamiltonian (\ref{Heff} ) may be characterized by a ``wave vector'' $k$ in the first Brillouin zone, which is equal to the charge $q$ modulo $2e$.  For energies less than $E_J$ we find a series of narrow tight binding bands,  of width $4\zeta$ , separated from each other by energy gaps  $ \approx \hbar \Omega_{JC}$.
For the lowest energy band we have 
\be
E_k \approx   -E_J + \frac{1}{2} \hbar \Omega_{JC} -2\zeta\cos (2\pi k/2e)
\label{qqeff}       
\ee
For energies above $E_J$, we find free running bands, separated by narrow energy gaps at the Brillouin zone boundaries $k = \pm e$ and at the zone center, $k=0$. 

If we now connect the junction to an ideal current source with current $I\neq 0$, we must take into account the term proportional to $I$ in (\ref{Heff}).
We thus obtain the Hamiltonian for a quantum particle in a tilted washboard potential, like that shown in Fig. (\ref{WB}a).

The current term acts like a force which changes continuously the
quasi-momentum $k$. If $I$ is not too large, a particle initially in a low energy state with $k \approx 0$  will accelerate until it
reaches the Brillouin zone edge, $k=e$ and then will be back-scattered
by a reciprocal lattice vector, into $k=-e$. (This is the origin of Bloch oscillations, where pulling on an electron in a periodic potential
produces an oscillatory motion back and forth, in the absence of dissipation.)   Physically, when  the charge on the capacitor plates
reaches $e$, a Cooper pair is transmitted through the junction and makes
the charge $-e$, and the process of charging repeats \cite{LikharevZorin,Averin,Haviland}. During this
process the phase $\phi$ oscillates  back and forth,  but there is {\it zero} average voltage
drop, while the time-average supercurrent is equal to the input current $I$.

There is,  however,  another type of energy eigenstate,  where the particle starts out with an energy above the top of the cosine potential and then accelerates to larger and larger velocities, with increasing value of $\phi$.  Moreover, a particle that is initially trapped in a low-energy Bloch-oscillation state will eventually tunnel out, by a series of Zener processes through higher tight binding bands into the runaway states.  The time scale for this will be very long, if $I$ is sufficiently small, but eventually it should happen, if the  system is truly described by the dissipationless washboard model. In the runaway state, the voltage steadily increases as the  charge $q$ builds  up continuously on the capacitor plates, while the time average supercurrent through the junction is zero. Thus the Josephson junction has become an insulator.

\subsection{Resistively shunted Josephson junction}

The case of a junction shunted only capacitatively is clearly a rather
pathological limit. In fact, 
the situation changes radically if we add a
shunt resistor in parallel to the junction. The resistor provides
a damping force which, depending on its size,  can either stabilize the system in state of localized $\phi$, where it carries a supercurrent with vanishing voltage drop, or can stabilize a modified form  of the runaway state, where $\phi$ increases linearly in time, giving rise to a finite voltage drop $V$, but with vanishing  time-average supercurrent.  In this latter case, the current $I$ is carried entirely by the shunt resistor, so that the Junction itself is essentially an insulator.

A systematic way of modeling the resistively-shunted Josephson junction (RSJ) is provided by the Caldeira-Legett model, to be described below.  However, we shall first
try to understand qualitatively
where a transition might occur between an insulating and
superconducting state in the limit of low currents. Essentially,
the phase of the junction is determined by the extent of uncertainty
in the phase of the junction. Since we are considering the limit of
low currents, let us ignore the current source altogether, and
consider what happens when a Cooper pair is transferred accross the
junction. The charge $2e$ creates a voltage imbalance which leads, in
turn, to a current through the resistor, until the imbalance is
relaxed. As current flows through the resistor,there is  a voltage drop
$V=I(t)R=\hbar\dot{\phi}/2e$ across the system, which causes  the phase
$\phi$ to wind. The total winding amount is:
\be
\Delta\phi=\int dt \dot{\phi}=\int dt 2e I(t)
R/\hbar=\frac{(2e)^2}{\hbar R}.
\ee
If the phase winding as a result of a Cooper-pair tunneling is higher
than $2\pi$, we expect that the phase coherence can not be maintained. Thus  we may 
conclude that for $R$ larger than some critical value, of order
\be
R_Q=\frac{h}{4e^2}=6.45k\Omega
\ee
the junction will be in its insulating phase in the limit of zero
current. The quantity $R_Q$  is
the ``quantum resistance" associated with Cooper pairs.
One can construct a dual argument and
caculate the amount of charge that gets transferred across the
resistor in case of a phase slip. This yields by an analogous
calculation $\Delta Q=2e R_Q/R$, and therefore indicates that for
$R<R_Q$ we can have a superconducting phase, since the charge
fluctuations are larger than a Cooper pair. Analysis of the Caldeira-Legget model indicates that in the limit of zero temperature and current, there is indeed a sharp transition between  superconducting and insulating states, and that this transition occurs precisely at $R=R_Q$.
This transition was originally predicted by Schmid
\cite{Schmid} (see also Refs.\cite{Bulgadaev,Korshunov1}) and for a related system by Chakravarty
\cite{Chakravarty}.

The 
 Caldeira-Leggett model,  for
$E_J\gg E_C$ is described by the imaginary time action:
\be
S_{RSJ}=\frac{1}{\hbar}T\summ_{\omega}\l[
  \frac{1}{2E_J}\frac{\hbar^2}{4e^2}|q_{\omega}|^2\omega^2+\frac{1}{2}|\omega||q_{\omega}|^2 R\rr]+\frac{1}{\hbar}\int d\tau \zeta \cos 2\pi\frac{q}{2e} ,
\label{CLaction}
\ee
where the sum is over Matsubara frequencies $\omega= 2 \pi n / T \hbar$  and the integral is over imaginary times $\tau$.  
The first term is the inductive energy in the Josephson junction due
to a current. The last term describes the hopping between two tight
binding states of the junction at $\phi=2\pi n$ and $\phi=2\pi(n\pm
1)$. The middle term is the Caldeira-Leggett term \cite{CaldeiraLeggett}, which imitates a term
describing the damping force due to a resistor. 
Caldeira and Leggett
 constructed the dissipative term by considering the Junction coupled to
 many oscillators, with a frequency distribution chosen to give the correct damping rate , and integrating out the oscillators.. 

The partition function of the RSJ, calculated as a path integral over
all imaginary time trajectories with a Boltzmann factor exponential in
the action (\ref{CLaction}), can be understood by expansion in
powers of $\zeta$. This yields the partition function of a
one-dimensional (imaginary time) gas of interacting phase slips, with
'charge' $\pm 1$ indicating the phase-slip's direction. The ``interaction" between 
phase-slips with charges $p_1$ and $p_2$ at times $\tau_1$ and $\tau_2$ give a contribution to the action of
\be
G{(\tau_1,\,\tau_2)}=2p_1p_2\frac{R_Q}{R}\log\l(\Omega|\tau_1-\tau_2|\r),
\label{qps6}
\ee

As we found previously for vortices in a film, the logarithmic interaction between phase
slips induces a phase transition between a superconducting state
where phase slips are bound in neutral pairs , and a resistive state with unpaired phase
slips. This transition, however, is a quantum transition between two
zero-temperature ground states. The depairing transition
occurs when the action to add an additional uncompensated phase slip
matches its 'quantum entropy' in imaginary time:
\be
\frac{R_Q}{R}\log \Omega L_T-\log \Omega L_T=0
\ee
Therefore when the shunt has $R>R_Q$, the junction itself is
insulating, and all current is forced to go through the shunt. Quite
generally, the interaction strength between two phase-slips is 
 $ 2R_Q$ divided by the combined shunt dissipation. This is a useful
principle for a quick analysis of quantum Josephson junction
systems. 

A combination of a  renormalization group analysis similar to that for the Kosterlitz-thouless transition  and heuristic
arguments provide us with the resistance of the RSJ as a function of
temperature. The phase slip fugacity renormalization is:
\be
\frac{d\zeta}{d\ell}=\l(1-\frac{R_Q}{R}\rr)\zeta
\ee
where the upper frequency cutoff is set at $  e^{-l} \Omega_{JC} $.
The initial value $\zeta_0$ ($\zeta$ at $l=0$) is given by Eq. (\ref{zetadef1}). 
If we carry out the RG flow until the point $l = l^*$, where  $2  e^{-l}  \Omega_{JC}  \sim T/ \hbar$, we can obtain an expression for
the resistance due to quantum phase slips. If $\zeta\rightarrow \Omega_{JC}$
during any stage of the RG flow, then superconductivity in the
junction breaks down. Otherwise, the probability rate of a phase-slip
occuring is:
\be
p_{\rm{ps}}\sim \l(\frac{\zeta_{l^*}    }{\Omega_{JC}}\r)^2
\label{qps10}
\ee
The rate $r$ of occurrence of phase slips of either sign, is given by the product of this probability and the renormalized frequency scale:
$ r=e^{-l^*}  \Omega_{JC}p_{\rm{ps}}$.
In the presence of a non-zero current $I$, the
potential drop $V$ is  determined by the difference of  the rates for forward and backward phase slip rates, which is a product of $r$ and the 
factor
$\sinh (h I / 2e T),
$
as in Eq. (\ref{vjj}), if we assume $ \hbar I \ll eT$.  These arguments lead
to a linear resistance of the Josephon junction (to be understood as
parallel to the shunt resistor)\cite{Korshunov2}:
\be
R(T)\sim R_Q \l(\zeta_{l^*}/\Omega_{JC}\r)^2\sim R_Q \l(\zeta_0/\Omega_{JC} \r)^2 \frac{1}{T^{2(1-R_Q/R)}}.
\label{RJJ}
\ee

Note that the qualitative behavior of a Josephson junction in the quantum regime depends crucially on the properties of the external circuit through the shunt resistance $R$.  In general, if the temperature is far below the energy gap of the superconductors on either side of the junction, there should be no contribution to the shunt conductance from tunneling of 
excited quasiparticles  across the junction.  This contrasts with the results in the classical regime, where the external circuit was found to influence the pre-exponential factor but not the activation energy for resistance in the junction. 
Experimentally, a superconductor to insulator quantum phase transition in a single Josephson junction tuned by resistance of the external circuit has been demonstrated in Ref. \cite{Penttila}. 

We have seen that even a small shunt resistance or dissipative coupling  can have major effects on the dc conductance of a Josephson junction in the quantum regime. However, in high-frequency  experiments, it may be possible to ignore dissipation, if the latter can be made sufficiently small. This is the driving principal in designs to use superconducting circuits as elements to construct a quantum computer\cite{Nakamura,Vion,Martinis,Chiorescu,Schreier,Hoskinson}. Although the general subject is outside the scope of this review, we mention one recent experiment where, after embedding a small Josephson junction in a superconducting circuit with high kinetic inductance, it was possible to observe coherent quantum tunneling between two adjacent wells of the $\cos \phi$ potential, with Rabi oscillations at a frequency 350 MHz.\cite{manucharyan} 
 (This is much smaller than the classical oscillation frequency within a well, $\Omega_{JC}/2\pi  \approx$ 13.5 GHz.)

Before concluding this section we would like to discuss
another perspective on the interplay of quantum fluctuations and
dissipation in Josephson junctions. Consider first the case of an
underdamped junction in the limit where $E_J \ll E_C$ which is
opposite to the regime we have been considering so far. Since
the shunt resistance is large compared to $R_Q$, the junction will be
in the usual Coulomb blockade regime, where there is an energy gap
$E_B\approx E_C$ for electrical transport. The vanishing of the linear
conductance of the Josephson junction in this regime appears quite
natural. RG analysis, however, predicts insulating behavior of Cooper
pairs for underdamped junctions even in the limit $E_J \gg E_C$, when
one would naively expect Coulomb blockade effects to be suppressed.  
The RG argument can be formulated as follows: in the
underdamped regime the probability of quantum phase slips
increases with lowering the temperature as $\sim
T^{-2(1-R_Q/R)}$. However the prefactor in this expression involves
$\zeta_0$, the probability of QPS at the microscopic scale
$\Omega_{JC}$ (see eq. (\ref{RJJ})). The latter is given by equation
(\ref{QPSaction}) and is exponentially small. Thus observing
insulating behavior of underdamped Josephson junctions in the regime
$E_J \gg E_C$ requires working at exponentially low temperatures and
currents\cite{LikharevZorinJLT,Penttila}.  We remark that non-linear
transport at non-zero voltages can be quite complicated in this regime
and we shall not attempt to discuss this here.  Results  depend on
many details of the environment. 

\cite{IngoldNazarov}.

In the discussion above  Ohmic dissipation was introduced in the form of a Caldeira-Leggett heat bath of harmonic oscillators.
This is the simplest quantum mechanical model which produces the correct classical equations of motion. One may also consider
more realistic microscopic models of dissipation, such as quasiparticle tunneling (see e.g. ref. \cite{Ambegaokar}). These models are more challenging
for theoretical analysis and result in a richer set of phenomena and more complicated phase diagrams (see e.g. Ref \cite{Guinea}).

\subsection{Quantum phase slips in wires: the quantum K-T transition \label{QKT}}

As we saw in Section (\ref{thermalwires}), thin superconducting wires, like mesoscopic Josephson junctions, will have a finite
phase-slip related resistance at any non-zero temperarture.  One may also ask, however, about the possibility of phase-slip events caused by quantum tunneling processes, which might be important at sufficiently low temperatures. According to our current theoretical understanding, as  discussed below, 
an infinitely long  wire of superconducting material can show  a phase transition at zero temperature, as a function of wire thickness, in which  superconductivity is destroyed by unbinding of phase slips in the space-time plane,  analogous to the finite-temperature Kosterlitz-Thouless transition in a two-dimensional film, or the zero-temperature phase transition in a single junction connected to a shunt resistor.

The simplest way to understand the phase-slip proliferation transition
in a wire is to think of it as a chain of superconducting
grains with self-capacitance, that are connected via Josephson
junctions. Each grain roughly represents a segment
of length $a\sim\xi(0)$ of the wire, and the Lagrangian describing the wire is then:
\be
\L_i=\frac{1}{2}C a
\l(\frac{\hbar}{2e}\der{\phi_i}{\tau}\rr)^2+\frac{J}{a}\cos\l(\phi_{i+1}-\phi_i\rr),
\label{wire-disc}
\ee
where $C$ is the capacitance per unit length, and $J$ is proportional to the one-dimensional superfluid density in the wire (We have assumed here that the capacitance to ground is more important than the capacitance between grains at the wavelengths  of the important fluctuations).  
The {\it effective}
impedance shunting a given Josephson junction is calculated by assuming that the other junctions are
perfectly superconducting, and therefore behave as inductors  for small fluctuations in the current. 
The effective impedance of two semi-infinite telegraph line (one on
each side of the junction) with per-length capacitance C and
inductance $\hbar^2/4e^2J$ is 
\be
Z=2\sqrt{\hbar/2eJC}.
\label{zeff}
\ee
As we discuss below,
the superconductor-insulator transition happens when $Z=R_Q/2$, where
the extra factor of 2 is due to the entropy arising from the spatial degree
of freedom of phase slips\cite{Haviland2}. 

A more quantitative analysis of the wire can continue along the lines
of the single junction analysis. The partition function of a wire, like a single junction, can be
written as that of a neutral gas of interacting phase slips in space-time with a logartihmic interaction\cite{Doniach,Fazio}. 
The strength of the interaction
is determined by the effective dissipation of the chain, given in
Eq. (\ref{zeff}), and for two phase slips  separated by a space-time vector 
$(x,\tau)$ it is:
\be
G{(x,\,\tau)}=p_1p_2 \sqrt{JC/4e^2} \log\l(\Omega\l(x^2/v_{MS}^2+\tau^2\r)^{1/2}\r).
\label{kt1}
\ee
where $v_{MS}^2=(4e^2/\hbar^2)J/C$ is the Mooij-Schoen mode: the speed with which phase
fluctuations propogate in the superconducting wires.
By thinking of the imaginary time direction $\tau$ as a second space
direction, we see that this Lagrangian coincides with the energy
density of films, Eq. (\ref{js}).  Phase slips are
clearly the space-time analog of vortices in 2d films.
Since we now have the restriction $|\tau| < T/\hbar$,  a quantum chain at finite $T$ corresponds to the classical behavior of a film of finite width.

If we use $y=v_{MS}\tau$, the dimensionless stiffness  $K$ of
the film is relaced by 
\be
K_Q = \sqrt{JC/4e^2}.
\label{KQ}
\ee
With this
classical-quantum mapping, we can infer all properties of the
wires.  We can use an RG analysis to describe the flow 
 of the plasmon-interaction
strength $K_Q$, and a phase-slip fugacity $\zeta$, as we  integrate out
modes of the phase $\phi$ with large frequencies and wave
vectors, and rescale both space and time.   Skipping technical details, the flow
equations one obtains are
\be
\ba{c}
\frac{dK_Q}{dl}=-\frac{\pi}{2}K_Q^2\zeta^2\vspace{2mm}\\
\frac{d\zeta}{dl}=\zeta\l(2-\frac{K_Q}{2}\r)
\ea
\label{kt3}
\ee

If we expand about the transition point $K_Q=4$, these equations have the same form as the Kosterllitz-Thouless flow equations. At $T=0$, if the initial value of $K_Q$ is sufficiently large, and $\zeta$ is small, one flows to a  point  on a ``fixed line", with $\zeta=0$ and $K_Q>4$.  This implies that the wire is a superconductor at $T=0$, with a renormalized value of  the superfluid density, or equivalently of $J$,   which is related to $K_Q$ by Eq. (\ref{KQ})
For temperatures that are non-zero but sufficiently small, one  finds a resistivity that decreases with $T$ according to the power law
\be
\rho(T)\sim T^{K_Q-3}.
\ee

\begin{center}
\begin{figure}
\includegraphics[width=10cm]{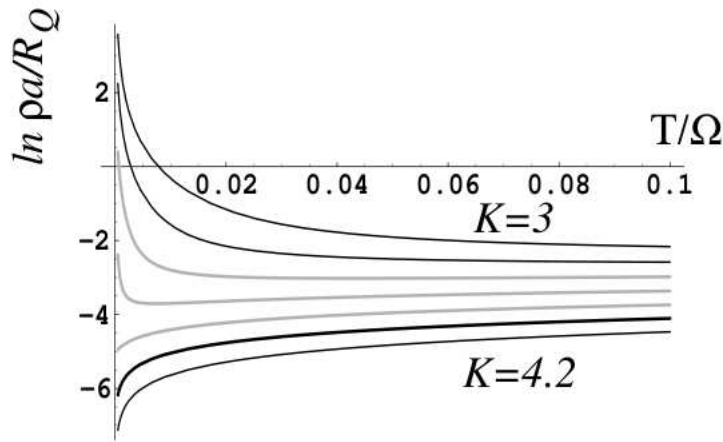}
\caption{Resistivity of a superconducting wire as a function of
  temperature in the vicinity of a Kosterlitz-Thouless
  zero-temperature phase slip unbinding transition. The parameter $K$
  takes the values $K=3,3.2,3.4,\ldots,4.2$ wiht $K=4$ being the
  quantum critical point. The grey lines correspond to the insulating
  phase of the wire, but, interestingly, they initially show a
  reduction in resistance as temperature decreases, and only at the
  lowest temperatures their resistivity curves up. In experiments,
  such effects will give rise to non-monotonic behavior of the
  resistivity. \label{KT-fig}}
\end{figure}
\end{center}

If at some point in the RG flow, the value of $K_Q$ becomes smaller than 4, the fugacity $\zeta$ will begin to increase, and $K_Q$ will then decrease to zero. (This can happen if the wire is too thin.)
The wire will then be an insulator at $T=0$.  Mirroring the behavior of $\zeta$, one predicts that for wires that are slightly on the insulating side of the transition,  the resistivity should first decrease  and then increase with decreasing temperature, eventually diverging as $T \to 0$.
Figure (\ref{KT-fig}) shows the traces of resistance vs. temperature
according to the K-T RG picture.

\subsection{Experiments on nanowires}

\begin{center}
\begin{figure}
\includegraphics[width=10cm]{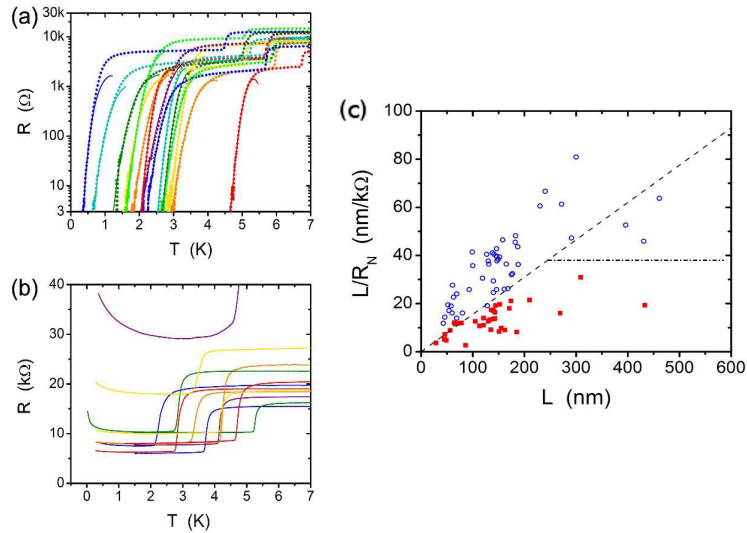}
\caption{MoGe nanowires experiments, taken from
  Ref. \cite{BollingerBezryadin}. (a) Resistance vs. temperature of
  superconducting samples. (b) Resistance  vs. temperature of
  insulating samples. (c) Phase diagram of all wires in a and b in
  terms of the normal state resistance (the resistance just after the
  leads turn superconducting, indicated by the 'knee' in the $R(T)$
  curves) and the conductivity. The dashed line corresponds to
  $R_N=R_Q$, and the dashed dotted line is added by here as the
  suggested long-wire critical conductance. \label{BezryadinFig}}
\end{figure}
\end{center}

Superconductivity and quantum phase slips in ultra thin quantum wires
was investigated in several experiments recently
\cite{AlbertChang,Tinkham, Bezryadin,BollingerBezryadin}. Particularly germane to
the discussion above were the nanowire experiments done by the Tinkham
and Bezryadin groups on MoGe amorphous nanowires (see ref. \cite{Bezryadin_review} for a review). These experiments
followed the resistance as a function of temperature for wires of
varius lengths ($100nm$ to $1\mu m$) and widths ($5nm$ to $25nm$). Figure \ref{BezryadinFig} summarizes some
of these experiments. 

The MoGe nanowire experiments clearly showed a transition between
weakly insulating behavior and superconducting behvior at low
temperature. Furthermore, the persistant resistance of the wires at
low temperatures indicated that this transition is driven by quantum
fluctuations. As Fig. \ref{BezryadinFig} shows, the location of the
transition is consistent with a global transition at $R_N=R_Q=6.5k\Omega$ for
short wires ($L<200nm$), where the wires behave like single shunted
junctions, and with a local infinite-wire like transition for longer
wires. This behavior is expected on the basis of phenomenological
two-fluid models \cite{RDOF1,RDOF2,Buechler}, and provides support to
the theory of quantum phase-slip proliferation. 

Nevertheless, a point of controversy is the detailed temperature dependence of the resistance
on the temperature. The measurements of Bezryadin, in
particular, show a rapid decay of resistance with decreasing
temperature for short superconducting nanowires\cite{Bezryadin_LAMH}, contrary to the
expectation of a power-law decay [Eq. (\ref{RJJ})]. In
Ref. \cite{MORS} it is shown that taking into account a finite density
of phase slips in short wires in a self consistent way, by modifying
the effective shunting resistance to be $R_{\rm {eff}}=R_N+\alpha\zeta^2$,
with $\zeta$ being the pahse slip fugacity, indeed produces sharp
declines of the resistance with decreasing temperatures in good
agreement with the experiments. 

While it is beyond the scope of this review, we would like to emphasize
that several other attmepts to describe the behavior of MoGe nanowires
were made. In particular, Refs. \cite{Sachdev stuff} showed that if
pair-breaking effects which give rise to dissipation in the nanowires
exist, one can describe the nanowires using the Hertz-Millis field
theory which gives rise to
a universal conductance at the superconductor-insulator transition
point. It was later shown by Vojta's group \cite{Hoyos} that any
amount of disorder would drive this field theory into an
infinite-randomness phase, which implies exotic scaling properties not
yet compared to experiment. Additionally, we must mention that quantum
effects are also expected to affect the mean-field transition
temperature (neglecting quantum phase slips)  of thin superconducting
wires. The theory for this suppression was developped by Finkel'stein
and Oreg, and shows good agreement with
experiment\cite{OregFinkelstein}.

\subsection{Quantum phase transitions in films \label{Qfilms}}

The study of superconductivity in thin films at low
temperatures is of particular contemporary interest. Quantum effects
may drive thin superconducting films into a resistive and even
insulating states at low temperatures. Similarly, disorder, which is always present in
experimental realizations of thin films, plays a crucial role in the
fate of superconductivity in films. 

Roughly speaking, films from superconducting materials undergo two classes of
superconducting-insulating transitions: magnetic field induced, and
disorer induced. In both cases the transition is between a
superconducting phase with a vanishing resistance at $T\rightarrow 0$,
and an insulating phase with a diverging resistance as $T\rightarrow
0$. Furthermore, a rough separation is made of 
films that undergo such a transition into two classes: granular and
amorphous films. We leave the review of the observed phenomena in
films to the review article by A.M. Goldman \cite{Goldmansreview}, also
included in this collection. Below we will briefly recount some of the
guiding principles in this topic.

The disorder effects on the superconducting state in amorphous films is observed
as a suppression of the critical temperature $T_c$ as the the thickness
of the film is reduced\cite{Beasley, Haviland,DynesValles}. By and large, this
phenomenon is explained by Finkelstein's analysis of the mean-field
transition in a thin diffusive metallic film
\cite{Finkelstein-films,OregFinkelstein}. The main idea is that Coloumb interactions suppress the transition to the
superconducting state more efficiently in diffusive thin films since
the time by which charge fluctuations can relax is shorter. For a
review see \cite{Finkelstein-films}.

In granular films, and in films in a finite magnetic field, it is expected
that Cooper pairs form, but fail to establish phase coherence
due to phase fluctuations induced by disorder and Coulomb
interactions \cite{Kapitulnik2007}. These phase fluctuations would give rise to a transition
between a superconducting state at low fields or when disorder is weak, and an insulating state at the opposite limits. Some
examples of disorder induced transitions are given in
Refs. \cite{Dynes}. The magnetic field induced
transitions occur in materials such as InO \cite{Shahar2004,
  Steiner2005}
and TiN \cite{Baturina2004,Baturina2008}, and produces insulating
states with a staggering resistance in
excess of $R_\square\sim 1G\Omega$. 

One illuminating, albeit only qualitative, picture for the quantum-fluctuations induced
transition is given in terms of vortices. A neutral gas of vortices describes quantum
fluctuations in the zero-field limit, and in a finite normal magnetic
field, there must be a net density of vortices.  A formal duality maps the
field theory of a bosonic gas (e.g., the Cooper pairs) to a field
theory of a gas of vortices, which are also considered bosonic \cite{FisherLee}. Since
the two theories are suspected to have similar universal properties
with regards to a formation of a condensed state, it also suggests
that at the superconducting transition the resistance
per square of the film should be of order $R_Q=h/4e^2$ (assuming a small Hall angle)\cite{Fisher1990}. 

To roughly see how it comes about, let us discretize the film into an array of
Josephson junctions.
Qualitatively, the film can be described {\it either} in terms of the
number of bosons (Cooper pairs) in each grain and their conjugate
phases ($n^{(CP)}_i,\,\phi_i$) {\it or} in terms of the number of vortices in each plaquette
and their conjugate phases ($n^{(V)}_i,\,\theta_i$). The transition
between the Cooper-pair superfluid and insulator is also a transition
between a localized-vortex phase and a vortex superfluid. At the
transition, both bosonic gases are diffusive, and their diffusion times should
also be similar. If we consider the resistance of the film, we can
also concentrate on a single representative bond (the 2d geometry
guarantees that this would also be the resistance per square). The
current across such a junction is $I=2e/\tau_{CP}$ with $\tau_{CP}$ the time for
a Cooper pair to cross the junction. Alternatively, in terms of vortex
motion, the voltage across the junction is
$V=\frac{\hbar}{2e}\frac{d\Delta\phi}{dt}\approx
\frac{\hbar}{2e}\frac{2\pi}{\tau_V}$, with $\tau_V$ the time constant
for vortex motion across the junction. If vortices and Cooper pairs
are close to being self dual at the transition, then we expect $\tau_V\sim \tau_{CP}$,
and therefore:
\be
R_c=\frac{V}{I}=\frac{\hbar}{2e}\frac{2\pi}{\tau_V}\frac{\tau_{CP}}{2e}\approx
\frac{h}{4e^2}.
\ee

Let us write the Cooper-pair Hamiltonian, and its
dual, the vortex Hamiltonian, explicitly. 
For simplicity, we
will assume there is no disorder in the discretized model, and that
the Cooper-pairs only experience short range repulsion. The Cooper-pair degrees of
freedom in a granular array have the Hamiltonian: 
\be
\H=-\summ_{<ij>}J \cos(\phi_i-\phi_j+\Phi_{ij})+U (n_i)^2
\label{cpa}
\ee
$J$ is the (nearest neighbor) Josephson coupling, $\Phi_{ij}$
accounts for the (physical) vector potential between the grains, $U$ is the
charging energy of each grain. This Hamiltonian can be recast in terms of the vortex density,
$n^{(V)}=\frac{1}{2\pi}\nabla\times\nabla\phi$, and an angle $\theta_i$ which is
conjugate to the vortex density. Note that in the vortex context the
index $i$ refers to plaquettes bounded by Josephson junctions. The
Hamiltonian is found to be
\be
\H_V  \approx-\summ_{<ij>}t^{(V)}\cos (\theta_i-\theta_j )+
U^{(V)}   \delta n^{(V)}_i  \delta n^{(V)}_j   \ln(r_i-r_j)   
  \label{va}
\ee
where $ \delta n^{(V)}_i \equiv n^{(V)}_i-H/\Phi_0$.
The first term in (\ref{va}) describes the hopping of vortices between adjacent
plaquettes, with hopping strength $t^{(V)}$; the stronger the charging
interactions (and, in principle, 
disorder) in the sample are, the larger is $t_V$. The
vortex-interaction parameter is roughly $U^{(V)}=\pi J$

The vortex Hamiltonian (\ref{va})  and the Cooper-pair Hamiltonian 
(\ref{cpa})  are both bosonic, but their details differ.
Thus the Cooper-pair - vortex duality is not an exact  
  self-duality. Nevertheless, it is thought that the two actions are
sufficiently similar that the resistance of films at
the superconductor insulator transition should be close to the value
$R_Q$ that an exact self-duality  would indicate.

The superfluid-insulator transition in 2d systems is still not fully
understood, and the vortex-Cooper-pair duality is so far a guiding
principle more than a theory. The experimental situation is made even
more complicated due to some samples exhibiting resistance saturation
in parameter regimes between the superconducting and insulating
regimes, as the temperature sinks below $T\sim 100mK$
\cite{MasonKapitulnik,Steiner2005,Yoon2006,Shahar2004}, which by some is
thought to be an intervening exotic metalic phase \cite{WuPhillips,
  vortexmetal} whose origins are unknown. While our discussion above was 
referring to superconducting films, it mostly applies to superfluid
Helium films as well. Like superconducting films, the appearance of
superfludity in Helium on strongly disordered substrate such as vycor
is still not fully clarified. (See, 
e.g., \cite{Chan1,Chan2}).

\section{ac conductivity}
\label{ac}

As was mentioned in the introduction, 
At non-zero temperatures, in the presence of a non-zero potential gradient,  there will be a contribution to the electrical current  from the motion of thermally excited quasiparticles. This normal fluid  contribution will be negligible compared to the supercurrent in a dc measurement, if the  phase slip rate $\eta$ is  sufficiently small,  since the potential gradient itself will be vanishingly small in this circumstance.  In an ac experiment, however, the  supercurrent  will be accompanied by a non-zero reactive voltage  even in the absence of vortex motion, and this voltage will lead to a non-zero normal current with associated dissipation.  

In the absence of vortex motion, we can write the current $\bj$ induced by a weak electric field $\bE$  at frequency $\omega$ as $\bj = \sigma(\omega) \bE$, where 
\be
\sigma(\omega) = \frac { i \rho_s}{m \omega } + \tilde{\sigma}_{n} (\omega)
\label{som}
\ee
where $\rho_s$ is the superfluid density, $m$ is the electron mass,  and $\tilde{\sigma}_{n} $ is the conductivity of the normal fluid, which approaches a non-zero real constant for frequencies below the scattering frequency of the quasiparticle excitations.  The superfluid density is related to the previously defined quantity $K$  by 
$\rho_s/m \equiv (2e)^2 K $.  The first term in (\ref{som}) defines a ``kinetic inductance" for the superconductor.

The combination of the Cooper-pair  inductance, and the normal
electrons' dissipation has important consequences for electronic
applications such as resonators and microwave cavities. The dissipated power per unit volume in a 
superconducting material with an ac current density $j$ is given by 
$\rho j^2$, where the ac resistivity  $\rho$  is defined by
$
\rho = {\rm Re}[ 1/\sigma(\omega)]  \approx  \omega^2 \tilde {\sigma}_{n} m^2 / \rho_s^2 .
$

In microwave cavities made of a
superconducting material, the normal fluid will be responsible for
losses on its surface. 
The skin depth  $\delta$ of the radiation  $\approx (mc^2/ 4 \pi \rho_s)^{1/2}$      
which is independent of $\omega$, and since   
since the total ac current per unit area of the surface for a given  intensity of the incident microwave power is also independent of $\omega$ , the ratio between the power absorbed in the surface and the incident power is proportional to $\omega^2 \tilde {\sigma}_{n}$.

In standard BCS superconductors, the minimum  energy to create a quasiparticle is given by  the energy gap $\Delta$.  The value of $\tilde{\sigma}_{n}$ is proportional to  the number of excited quasiparticles, which goes to zero as $e^{-\Delta/T}$ at low temperatures.  However, one can also have superconductivity without an energy gap, as in the presence of magnetic impurities.  In such cases, the normal fluid conductance will not vanish exponentially at low temperatures, and a more complicated analysis is necessary. See, e.g., Refs.
\cite{Maki-parks,DeGennes}. Of course for frequencies larger than $2 \Delta/ \hbar$ there would be power absorption even at $T=0$.

\section{ Systems of Ultra-Cold Atoms}
\label{cold-atoms}

Recent experiments with  ultracold alkali atoms have opened a new chapter
in the study of nonequilibrium dynamics of superfluids. Two features
of such systems make them particularly suitable for studying
dynamical phenomena: complete isolation from the environment and
characteristic frequencies of the order of kHz, which are readily
accessible to experimental analysis. Dynamics of atoms in optical
lattices \cite{jaksch} has particularly close connection to the
issues that we discussed earlier in the context of superconductors.
Here we restrict our discussion to
bosonic atoms although interesting experiments have also been done
with fermions (see e.g. Ref. \cite{miller}).

Optical lattices are created using standing waves of laser beams that
provide an artificial periodic potential for the atoms. The strength
of the optical potential can be controlled and atomic systems can be
tuned between the regimes of a  weak lattice, where kinetic energy
dominates, and strong lattice, where repulsive interactions between
atoms play the dominant role\cite{jaksch,fisher}. In the former case
one finds BEC of weakly interacting atoms and a macroscopic
occupation of the state with quasimomentum $k=0$. In the latter case
the system is in the Mott insulating state which has no long range
phase coherence and strongly reduced number fluctuations. The transition
between the two phases is an example of the quantum fluctuations
driven phase transition which we discussed in Sections \ref{QKT} and \ref{Qfilms}. Such
a transition was first observed by Greiner et al. \cite{greiner} by
measuring momentum occupation numbers in the so-called
time-of-flight experiments.

Experimentally, one can also prepare cold atoms systems moving with
respect to the optical lattice and study the decay of the current.
Such experiments are very similar in spirit to the critical current
measurements in the case of superconductors. In the weakly
interacting regime one expects the critical current to be determined by
the inflection point of the single particle
dispersion\cite{niu,smerzi,fallani}. Beyond the inflection point the
effective mass becomes negative, which is equivalent to a change of
interaction from repulsive to attractive, so small density
fluctuations become amplified making the system unstable to
fragmentation. Close to the Superfluid to Mott transition we expect
the critical current to go to zero continuously. So the question is
how to connect the two regimes. This was analyzed both theoretically
\cite{altman} and experimentally\cite{fertig,mun,demarco}, and
direct signatures of both thermal and quantum phase slips have been
observed.

\section{Conclusion}
We have seen that the mechanisms for production of resistance in superconducting materials are understood in broad outline.  However, there remain many open questions, particularly at very low temperatures, when quantum fluctuations are important. In such situations, the dynamics of phases slips or vortex motion will be sensitive to couplings to sources of dissipation, internal or external, possibly at multiple frequencies, and we have only a limited understanding of how this occurs in  actual experiments. Open questions exist even for  presumably classical problems, such as  the resistance  of a superconducting wire close to near $T_c$, where the simple time-dependent 
Ginzburg-Landu model  seems to work much better than it should.  
   
There are many open questions regarding the role of disorder in the classical as well as quantum regimes.  We understand only partially the collective pinning that results from the interplay of disorder and  vortex-vortex repulsion  for a type-II superconductor in a strong magnetic field.  Issues of how to increase pinning and decrease flux creep are of great importance for practical applications of superconductors in the areas of power transmission and high field magnets.  

Review articles and books, as well as articles in the original literature,  point to  open issues in the field.  Among the helpful examples are  Refs  \cite{Schoen-Zaikin,IngoldNazarov, Likharev,Arutyunov,Simanek,Fazio,Bezryadin_review}

\subsection*{Acknowledgments}
The authors have benefited from  discussions  with many people on topics of this review,   including, in recent years,  M. Tinkham, Y. Oreg, A. Bezryadin, I. Aleiner, D. S. Fisher, M. P. A. Fisher, 
V. Galitskii, and C. L. Lobb.  They acknowledge support from the Packard Foundation, a Cottrell Fellowship from the Research
Corporation, and  NSF grants DMR- 0906475 and DMR-0705472.

\bibliographystyle{ws-rv-van}
\bibliography{bib1}

\printindex                         
\end{document}